# Coexistence of two superconducting phases in $Ca_{1-x}La_xFe_2As_2$

Y. Sun, W. Zhou, L. J. Cui, J. C. Zhuang, Y. Ding, F. F. Yuan, J. Bai, Z. X. Shi[*]

*Department of Physics and Key Laboratory of MEMS of the Ministry of Education, Southeast University, Nanjing 211189, People's Republic of China*

*e-mail address:* zxshi@seu.edu.cn

## *Abstract*

Single crystals of $Ca_{1-x}La_xFe_2As_2$ for $x$ ranging from 0 to 0.25 have been grown and characterized by structural, transport and magnetic measurements. Coexistence of two superconducting phases is observed, in which the low superconducting transition temperature ($T_c$) phase has $T_c \sim 20$ K, and the high $T_c$ phase has $T_c$ higher than 40 K. These data also delineate an $x$-$T$ phase diagram in which the single magnetic/structural phase transition in undoped $CaFe_2As_2$ appears to split into two distinct phase transitions, both of which are suppressed with increasing La substitution. Superconductivity emerges when $x$ is about 0.06 and coexists with the structural/magnetic transition until $x$ is $\sim 0.13$. With increasing concentration of La, the structural/magnetic transition is totally suppressed, and $T_c$ reaches its maximum value of about 45 K for $0.15 \leq x \leq 0.19$. A domelike superconducting region is not observed in the phase diagram, however, because no obvious over-doping region can be found. Two superconducting phases coexist in the $x$-$T$ phase diagram of $Ca_{1-x}La_xFe_2As_2$. The formation of the two separate phases, as well as the origin of the high $T_c$ in $Ca_{1-x}La_xFe_2As_2$ is studied and discussed in detail.

## 1. INTRODUCTION

The recent discovery of superconductivity at 26 K in the iron oxypnictide LaFeAs(O,F)[1] has stimulated great interest in the condensed-matter physics community. A tremendous amount of work has been carried out, leading to the emergence of novel iron-based superconductor families with different crystal structures: 1111 (LaFeAs(O,F)),[1] 122 ((Ba,K)Fe$_2$As$_2$),[2] 111 (LiFeAs)[3] and 11 (Fe(Se,Te))[4]. These compounds adopt a layered structure, based on FeAs(Se) layers, from which the superconducting carriers mainly flow. Similarly to the cuprate superconductors,[5] superconductivity in iron oxypnictide appears to be accompanied by the suppression of an antiferromagnetic (AFM) state upon doping or applying pressure, leading to very similar phase diagram. In cuprate and oxypnictide superconductors, electron and hole doping can both successfully induce superconductivity, although the symmetry that appears between electron- and hole-doping in cuprate superconductors[6] is still not clear in oxypnictide superconductors. In the parent LaFeAsO, superconductivity can be induced by both F$^-$ and Sr$^{2+}$ doping in O and La-sites, respectively, the results of which proved to be electron and hole type carriers. The highest superconducting transition temperature, $T_c$, in the two compounds are both about 25 K, and the phase diagrams are similar.[7] On the other hand, in the 122 system, hole-doped Ba$_{1-x}$K$_x$Fe$_2$As$_2$[2] was reported to attain superconductivity at 38 K, while electron-doping can only induce superconductivity at a lower temperature, around 22 K.[8] One possible reason is that the electron-doping in the 122 system is usually realized by substituting other transition metals for the Fe ions in the FeAs layer, which is believed to be the main carrier conducting layer. Thus, an electron-doped 122 superconductor with perfect FeAs layers will be an ideal candidate to study the symmetry between electron-doping and hole-doping. Recently, superconductivity over 40 K was discovered in rare-earth doped CaFe$_2$As$_2$,[9-14] which has already proved to be dominated by electron-like charge carriers. Although these previous reports have revealed some of the structure, magnetic and transport properties of this compound, the evolution of the structural/magnetic transition and superconductivity with rare earth doping is still unknown. Furthermore, in order to understand the conditions for superconductivity and probe the symmetry between electron- and hole-doping in the 122 system, temperature versus doping phase diagrams must first be constructed, which are still not clear, unfortunately in Ca$_{1-x}$RE$_x$Fe$_2$As$_2$, where RE is a rare earth element. In this article, we report a systematic investigation of the phase diagram of Ca$_{1-x}$La$_x$Fe$_2$As$_2$ single crystals for $x$ ranging from 0 to 0.25 by using structure, transport and magnetic measurements. The origin as well as the characteristics of the high transition temperature of $Ca_{1-x}RE_xFe_2As_2$ is also studied and discussed in detail.

## 2. EXPERIMENTAL METHODS

Single crystals of $Ca_{1-x}La_xFe_2As_2$ with $x$ ranging from 0 to 0.5 were grown using the FeAs self-flux method. The FeAs precursor was first synthesized by reacting stoichiometric amounts of Fe and As inside a vacuum quartz tube at 750 ℃ for 24 h. High purity Ca grains, La bulks and FeAs powders, mixed together in the ratio 1-$x$: $x$: 4, were put in alumina crucibles and sealed in a quartz tube under a 30% partial pressure atmosphere of Ar gas. The sealed quartz tube was quickly heated to 1180 ℃, and kept at this temperature for 2 h, then slowly cooled down to 970 ℃ at a rate of 2 ℃/hour. After that, the temperature was cooled down to room temperature by shutting down the furnace. Single crystals with a typical size of 5×5×0.2 mm were easily obtained by mechanically cleaving them from the flux. Single crystals were characterized by X-ray diffraction (XRD), with Cu Kα radiation from 10º to 70º. The actual La concentration was determined by a scanning electron microscope (SEM, Quanta 200) equipped with an energy dispersive X-ray spectroscope (EDX). Longitudinal and transverse (Hall) resistivities was measured by using a physical properties measurement system (PPMS, Quantum Design), and the magnetic susceptibility was measured by a Quantum Design superconducting quantum interference device ─ magnetic properties measurement system (SQUID-MPMS).

## 3. RESULTS AND DISCUSSION

Figure 1 presents the actual La doping $x$ measured by EDX versus the nominal concentration. The actual values of $x$ are almost the same as the nominal compositions at lower doping level, and they linearly increase until $x \sim 0.2$. Then, the actual concentration $x$ becomes smaller than the nominal one and saturates to the actual value of 0.25. This behavior is similar to that obtained by wavelength-dispersive spectroscopy (WDS),[9] as well as the EDX results[13] from other reports. Figure 2 shows the single-crystal XRD patterns for $Ca_{1-x}La_xFe_2As_2$. Only the (00$l$) peaks were observed, suggesting that the crystallographic $c$ axis is perfectly perpendicular to the plane of the single crystal. The lattice constant $c$ has been calculated and is plotted in the inset of Figure 2. There is no obvious change in the lattice constant during the La doping, which may be due to the close match between the ionic radii of La (130 pm) and Ca (126 pm).[9, 13]

The temperature dependence of the in-plane resistivity $\rho(T)$ of $Ca_{1-x}La_xFe_2As_2$ single crystals is shown in Figure 3 (a). The data for each sample are normalized by the room-temperature value $\rho(300K)$, and subsequently shifted by 0.3 on the y-axis for clarity. Resistivity of the undoped parent compound $CaFe_2As_2$ exhibits metallic behavior over the entire temperature range with a sharp step-like increase at the temperature of about 160 K. The anomaly in resistivity is associated with the structural, $T_s$, and magnetic, $T_M$, phase transitions.[15] With La doping, the anomaly gradually broadens and shifts to lower temperature, and it disappears when the proportion of La is over 13%. The suppression of the resistivity anomaly can also be seen clearly in the $d(\rho(T)/\rho(300K))/dT$ curve for $x = 0.06$ shown in Figure 3(b). No distinct difference between the structural and magnetic transitions is manifested for the $x = 0$, and 0.04 samples (for clarity, just the data for $x = 0$ is shown), although, with the increasing La doping, the combined structural/magnetic transition splits into two anomalies. Although no detailed study on the thermodynamic or transport properties has been conducted to distinguish these two phase transitions in $Ca_{1-x}RE_xFe_2As_2$, based on the results on the analogous Ba-122 compounds[16, 17] and Co-doped Ca-122,[18] it is natural to attribute the higher temperature to the structural phase transition and the lower temperature to the magnetic phase transition. The criteria of $T_s$ and $T_M$ are as usually used in similar compounds.[18-21]

Along with the suppression of the anomaly, superconductivity emerges when $x$ is about 0.06, and it coexists with the structure/magnetic transition. This resistivity behavior for La doping in the low doping region is very close to that of $Ca(Fe_{1-x}Co_x)_2As_2$.[18] With more La doping, however, $T_c$ is not totally suppressed after it reaches its maximum value, which is quite different from the case of Co-doped $CaFe_2As_2$ as well as other electron or hole doped iron-based 122 samples, in which a whole superconductivity dome was detected.[18,20-22]

Another distinct feature of the temperature dependent resistivity of $Ca_{1-x}La_xFe_2As_2$ is the two superconducting transition steps, which are clearly shown in Figure 3(c). The feature of two superconducting phases is also observed in Pr-doped $CaFe_2As_2$.[11] Here we found that the two phases feature exists at the medium doping level from $x = 0.13$ to 0.19 as shown in Figure 3 (c). To further separate the two phases and accurately obtain the $T_c$ of each transition, the derivation of the temperature dependent resistivity $d(\rho/\rho_{50K})/dT$ versus $T$, was plotted in Figure 3(d). For clarity, just the data for $x = 0.15$ and $0.17$ are shown, together with selected data for comparison from the two $x$ values on either side, where the two phase feature was not observed. The transition temperatures were obtained as the beginnings of the peaks in the $d(\rho/\rho_{50K})/dT$ curves. $T_{cH}$ and $T_{cL}$ are defined as the higher and lower transition temperature respectively. Thus, $Ca_{1-x}La_xFe_2As_2$ can be divided into three doping regions: the low doping region with only one $T_c$, which is lower than 20 K; the medium doping region with two separate $T_c$s, where the lower one, $T_{cL}$, is about 20 K, and the higher one, $T_{cH}$, is over 40 K; and the high doping region with only one $T_c$ at about 40 K.

To further confirm the transition temperature of the samples, the temperature dependence of magnetization, $M$-$T$, was measured on several selected samples from the three different doping regions mentioned above, and the results are shown in Figure 4, together with the temperature dependence of the resistivity for comparison. In the low doping region, the transition temperature obtained from the magnetic susceptibility measurement, $T_{cM}$, is almost the same as that from the transport results. In the medium doping region, $T_{cM}$ is close to the lower transport value, $T_{cL}$. In the high doping region, however, $T_{cM}$ is only around 20 K, much lower than the value of about 40 K obtained from the resistivity measurement and close to the lower $T_c$ of the medium doping region. Compared to the transport measurements, which only probe the superconducting percolative paths, the magnetic susceptibility results show the bulk properties of the sample. Thus, the lack of detection of the high $T_c$ phase in $M$-$T$ demonstrates that the superconducting phase with $T_c$ higher than 40 K is not a bulk property. The much lower $T_c$ observed from magnetic susceptibility than from the transport measurements has also been reported previously.[9, 10, 12] Although a $T_c$ higher than 40 K from $M$-$T$ was reported,[11] it is easily suppressed by a very small field of about 100 Oe. The lower $T_c$ phase was further detected by magnetic susceptibility measurements with different fields on the $Ca_{0.83}La_{0.17}Fe_2As_2$ sample, as shown in the inset of Figure 4(b), which demonstrates that the $T_c$ remains constant with increasing field. Thus, the low $T_c$ phase seems to be a robust global SC phase. The value of $T_c \sim 20$ K is also observed in La doped $SrFe_2As_2$,[23] and it is close to the values in other electron doped 122 samples such as $Ca(Fe_{1-x}Co_x)_2As_2$,[18] and $Ba(Fe_{1-x}M_x)_2As_2$.[20,21] On the other hand, the high $T_c$ above 40 K in $Ca_{1-x}RE_xFe_2As_2$ may be coming from filamentary or interface superconductivity, as will be discussed later.

Now, we must reconsider the three different doping regions identified above in the discussion of resistivity, and do a short summary. In the low doping region, just one low $T_c$ phase is observed. Then, with increasing La doping, the high $T_c$ phase emerges. In this medium doping region, the volume of high $T_c$ phase is still very low, and it is in disjoin patches separated by the low $T_c$ phase, so both phases can be detected by the transport measurements. Finally, in the high doping region, the volume of high $T_c$ phase is enough to form a continuous percolative path, so that the current will no longer pass through the low $T_c$ part of the sample, and it cannot be detected by transport measurements. Even though the high $T_c$ phase can form a continuous percolative path, its volume with respect to the whole sample is still too small to be observed in the magnetic susceptibility measurements.

Based on the resistivity and magnetic measurements described above, we can establish a doping-temperature ($x$-$T$) phase diagram for $Ca_{1-x}La_xFe_2As_2$, as shown in Figure 5. On the under-doped side of the phase diagram ($x \leq 0.13$), the structural/magnetic phase transition is monotonically suppressed with increasing La substitution, while at the same time, superconductivity emerges from $x = 0.06$, and then coexists with the orthorhombic/antiferromagnetic phase until $x = 0.13$, when the structural/antiferromagnetic phase transition is totally suppressed. The evolution of structural, magnetic and SC phases with La doping in the under-doped region is similar to what occurs in other electron-doped iron-based 122 superconductors. The suppression rate of $T_s/T_w$, however, is roughly 6 K per atomic percent La substitution much smaller than the value of 15 K per atomic percent Co doping in $Ba(Fe_{1-x}Co_x)_2As_2$ and 10 K for $Ca(Fe_{1-x}Co_x)_2As_2$.[18, 21] The structural/magnetic phase transition can also persist to 13% La substitution, which is higher than 5.8% for $Ba(Fe_{1-x}Co_x)_2As_2$ and 7.5% for $Ca(Fe_{1-x}Co_x)_2As_2$.[18, 21]

We now focus on the SC region of the phase diagram, which contains two separate $T_c$ phases. As discussed above, the lower value of $T_{cL}$ obtained from $R$-$T$ measurements together with the value of $T_{cM}$ from $M$-$T$ measurements corresponds to the low $T_c$ phase. Upon the substitution with La, the low $T_c$ phase keeps its $T_c$ value of roughly 20 K. Meanwhile, the high $T_c$ phase emerges, and the value of $T_c$, which has been increasing with La doping, reaches its maximum value of about 45 K for $0.15 \leq x \leq 0.19$, and then decreases to a lower value around 30 K. The high $T_c$ phase shows an incomplete domelike appearance, which does not reach the over-doping region, as $T_c$ cannot be totally suppressed. Here, we must emphasize that because of its non-bulk nature, the absolute $T_c$ value of the high $T_c$ phase may have some fluctuations between different pieces. The two-phase property is independent of the crystal, however, and it will not affect the discussion on the origins of the high $T_c$ phase below.

Before probing the origins of the two separate SC phases, the doping type of $Ca_{1-x}RE_xFe_2As_2$ should be first considered. Figure 6 shows the temperature dependence of the Hall coefficient $R_H$ of $Ca_{0.95}La_{0.05}Fe_2As_2$ and $Ca_{0.79}La_{0.21}Fe_2As_2$. For $Ca_{0.95}La_{0.05}Fe_2As_2$, the sharp increase in the absolute value of $R_H$ below $T_M$ indicates a sudden drop in the carrier density with the magnetic transition, which is a common feature in iron-based superconductors.[24] In the case of $Ca_{0.79}La_{0.21}Fe_2As_2$, the magnetic transition has been totally suppressed, and the strong temperature dependence of $R_H$ is often attributed to the effects of multiband or the pseudogap.[25, 26] The negative values of $R_H$ for both samples indicate that the dominant carrier for $Ca_{1-x}RE_xFe_2As_2$ is the electron. By using $R_H = 1/ne$, the charge carrier density, $n$, was roughly estimated and is shown in the inset of Figure 6. Compared to $Ca_{0.95}La_{0.05}Fe_2As_2$, the charge carrier density in $Ca_{0.79}La_{0.21}Fe_2As_2$ is obviously enhanced, since more La doping induces more electrons into the system.

The origin of the observed high $T_c$ phase in $Ca_{1-x}La_xFe_2As_2$ may be explained by four possible scenarios (*i*) Josephson junction coupling across the grains; (*ii*) filamentary superconductivity caused by local strength or defects; (*iii*) minor foreign phase; and (*iv*) interface superconductivity. Scenario (*i*) usually accounts for the observation of two superconducting transitions in granular polycrystalline superconductors, which violates the good crystallinity of the single crystal observed from the XRD results. In the favor of scenario (*ii*), filamentary superconductivity is a common problem in 122 iron-based superconductors, and has already been reported in $BaFe_2As_2$, $SrFe_2As_2$ and $CaFe_2As_2$.[27-29] In this case, some local or mesoscopic structural defect or the surface strain similar to that in some thin films will cause some strength, which will induce superconductivity in a very small fraction of the sample. The filamentary superconductivity caused by local strength or defects is usually sensitive to or even easily removed by heat-treatment, pressure, or magnetic field.[27-31]

Annealing at high temperature is an easy and direct way, proved effective in $SrFe_2As_2$ [29] and $CaFe_2As_2$ [28] to eliminate the filamentary superconductivity. Thus, we vacuum annealed the $Ca_{1-x}La_xFe_2As_2$ single crystal at temperatures ranging from 300 ℃ to 800 ℃ from several hours to as long as two weeks. Figure 7 shows typical resistivity results for $Ca_{0.79}La_{0.21}Fe_2As_2$ annealed at 800 ℃ for two weeks. Although the residual resistivity ratio (RRR) increased a little due to the removal of some defects, $T_c$ didn't change after annealing, as can be seen more clearly from the enlarged transition part in the inset of Figure 7. Recently Swee K. Goh[32] reported a pressure dependent resistivity measurement on $Ca_{1-x}La_xFe_2As_2$, which showed that $T_c$ is not suppressed, even under pressure greater than 40 kbar. Actually, an obvious two step resistivity transition was also witnessed in their sample.

To check the influence of magnetic field on the high $T_c$ phase, the temperature dependence of the resistivity with applied fields up to 9 T was measured on $Ca_{0.9}La_{0.1}Fe_2As_2$, $Ca_{0.83}La_{0.17}Fe_2As_2$, and $Ca_{0.79}La_{0.21}Fe_2As_2$ (selected from the three different doping regions of the phase diagram) and is plotted in Figure 8(a-c). Magnetic fields were applied along the $c$-axis, which has already been proved to be more sensitive to the field than the $ab$ plane.[10, 12] $T_c$ is gradually suppressed to lower temperature, and the transition is broadened with increasing magnetic field, which is similar to what has been reported for 1111 phase.[33, 34] The upper critical fields, $H_{c2}$, defined by 90 % of normal state resistivity are plotted in Figure 8(d) for three different $x$ values. For $Ca_{0.79}La_{0.21}Fe_2As_2$, with only the high $T_c$ phase, the slope of $H_{c2}$ is evaluated as -13.8 kOe/K from the linear fit to the $H_{c2} - T$ curve above 2 T. Using the Werthamer-Helfand-Hoenberg formula,[35] $H_{c2}(0) = -0.693 T_c dH_{c2}/dT \big|_{T=T_c}$, $H_{c2}^{0.21}(0)$ is simply estimated as 38.5 T. In the case of $Ca_{0.83}La_{0.17}Fe_2As_2$, two upper critical fields were obtained, belonging to the high $T_c$ phase and the low $T_c$ phase respectively, which again proves the existence of two phases in the medium doping region of $Ca_{1-x}RE_xFe_2As_2$. With the slope of $H_{c2}$ about -7.7 kOe/K and -11.0 kOe/K, the upper critical fields at 0 K for the high $T_c$ and low $T_c$ phases are 22.4 T and 13.7 T, respectively. The $H_{c2}$ at 0 K for $Ca_{0.9}La_{0.1}Fe_2As_2$ can be obtained by the same method as 6.8 T, although it shows a positive curvature different from the negative one in $Ca_{0.83}La_{0.17}Fe_2As_2$ and $Ca_{0.79}La_{0.21}Fe_2As_2$, which may come from the inhomogeneity in the low doping level of La. A detailed study of the upper critical fields, as well as the resistivity tail, which is a common feature of the $Ca_{1-x}RE_xFe_2As_2$,[9-12] was reported in another report.[36] Here, the results on the field dependent resistivity demonstrate that the two phases are not fragile with respect to the magnetic field. Thus, the heat treatment, field, and pressure dependent superconducting transition temperature results indicate that the high $T_c$ phase is not a filamentary type superconductivity caused by local pinning strength or defects.

For scenario (*iii*), although powder XRD patterns (data not shown) demonstrate that the diffraction peaks can be well indexed by the structure of $Ca_{1-x}La_xFe_2As_2$, and only impurity peaks from FeAs flux can be observed, we cannot simply exclude the possibility of the existence of trace amounts of foreign phase which cannot be distinguished by XRD. The minor foreign phase that might possibly exist is not a polycrystal-like impurity phase, though, because the upper critical fields[12,14] show anisotropy of about 3. It may be caused by the chemical phase separation. With increasing La doping, small amounts of foreign phase with high $T_c$ will emerge and coexist with the low $T_c$ phase. In this case, two superconducting phases would be observed. Then, as La doping is further increased, the high $T_c$ phase can form a continuous percolative path in the *ab* plane, and thus just one phase can be observed in the high doping region. With regards to scenario (*iv*), interface superconductivity can have an enhanced $T_c$.[37] In the case of $Ca_{1-x}La_xFe_2As_2$, the interfaces may be provided by alternate stacking of perfect and defective FeAs layers. Further work, especially on the microstructure/composition, will hopefully allow us to distinguish the origin of the high $T_c$ phase in $Ca_{1-x}RE_xFe_2As_2$ between the minor foreign phase and the interface superconductivity

## 4. CONCLUSION

In summary, we have successfully grown single crystals of $Ca_{1-x}La_xFe_2As_2$ ( $0 \leq x \leq 0.25$ ), and determined the phase diagram based on transport and magnetic measurements. We find that the single magnetic/structural phase transition for $CaFe_2As_2$ splits with La substitution. The superconductivity emerges when $x$ is about 0.06 and coexists with the structural/magnetic transition until $x \sim 0.13$, when $T_c$ reaches its maximum value of about 45 K. We did not observe a domelike SC region, however, because La substitution seems to saturate at $x \sim 0.25$, and cannot reach the over-doping region. Another distinct feature of the phase diagram is the two separate SC phases, of which the low $T_c$ phase is a robust global superconductivity and close to the SC phases of other electron-doped iron-based 122 samples. On the other hand, the non-bulk high $T_c$ phase with $T_c$ higher than 40 K may be caused by a minor foreign phase or interface superconductivity.

## 5. Acknowledgments

This work was supported by the Natural Science Foundation of China, the Ministry of Science and Technology of China (973 project: No. 2011CBA00105), the Scientific Research Foundation of the Graduate School (Grant No. YBJJ1104) of Southeast University, the Scientific Innovation Research

Foundation of College Graduates in Jiangsu Province (CXZZ_0135), and Jiangsu Science and Technology Support Project (Grant No. BE2011027).

# Figure captions

Figure 1: EDX result for the actual La concentration vs nominal La concentration for $Ca_{1-x}La_xFe_2As_2$.

Figure 2: Single crystal X-ray diffraction pattern of $Ca_{1-x}La_xFe_2As_2$. The inset shows the change of lattice constant $c$ vs the concentration of La.

Figure 3: (a) Temperature dependence of the in-plane resistivity for $Ca_{1-x}La_xFe_2As_2$, normalized to the room temperature value. Each subsequent curve for the next $x$ value is shifted downward by 0.3 for clarity. (b) $d(\rho(T)/\rho(300K))/dT$ for $x = 0$ and 0.06. The arrows indicate the magnetic ($T_m$) and the structural transitions ($T_s$). (c) $\rho/\rho_{50K}$ for $0.13 \le x \le 0.19$, where the two steps of superconducting transition can be clearly distinguished. (d) $d(\rho/\rho_{50K})/dT$ for $x = 0.1, 0.15, 0.17$ and 0.21. The arrows show the determination of $T_c$ for the two superconducting phases.

Figure 4: (a) *R-T* and (b) *M-T* results of $Ca_{1-x}La_xFe_2As_2$. The inset of (b) shows *M-T* measurements with different fields on the $Ca_{0.83}La_{0.17}Fe_2As_2$ sample.

Figure 5: *T-x* phase diagram of $Ca_{1-x}La_xFe_2As_2$ obtained from magnetic and transport results.

Figure 6: Temperature dependence of the Hall coefficient $R_H$ for $x = 0.05$ and 0.21. Inset shows the temperature dependence of the charge-carrier density $n$ for these two samples.

Figure 7: Temperature dependence of resistivity of $Ca_{0.79}La_{0.21}Fe_2As_2$ before and after vacuum annealing at 800 ℃ for two weeks. Inset is an enlargement of the area containing the $T_c$.

Figure 8: (a-c) Temperature dependence of resistivity with applied fields up to 9 T measured on samples $x = 0.1, 0.17$, and 0.21 (selected from the three different doping regions of the phase diagram). (d) Upper critical fields of samples $x = 0.1, 0.17$ and 0.21.

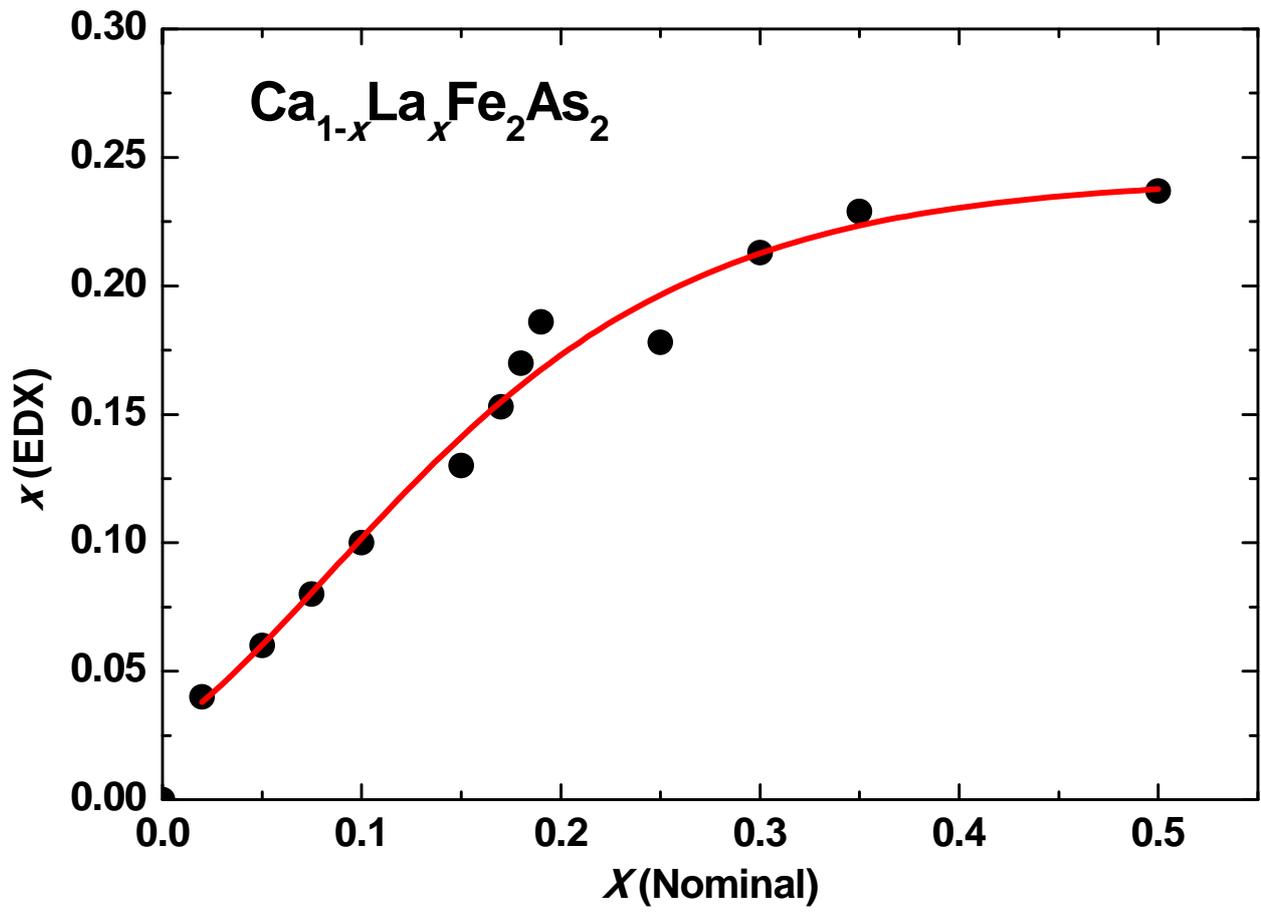

Figure 1

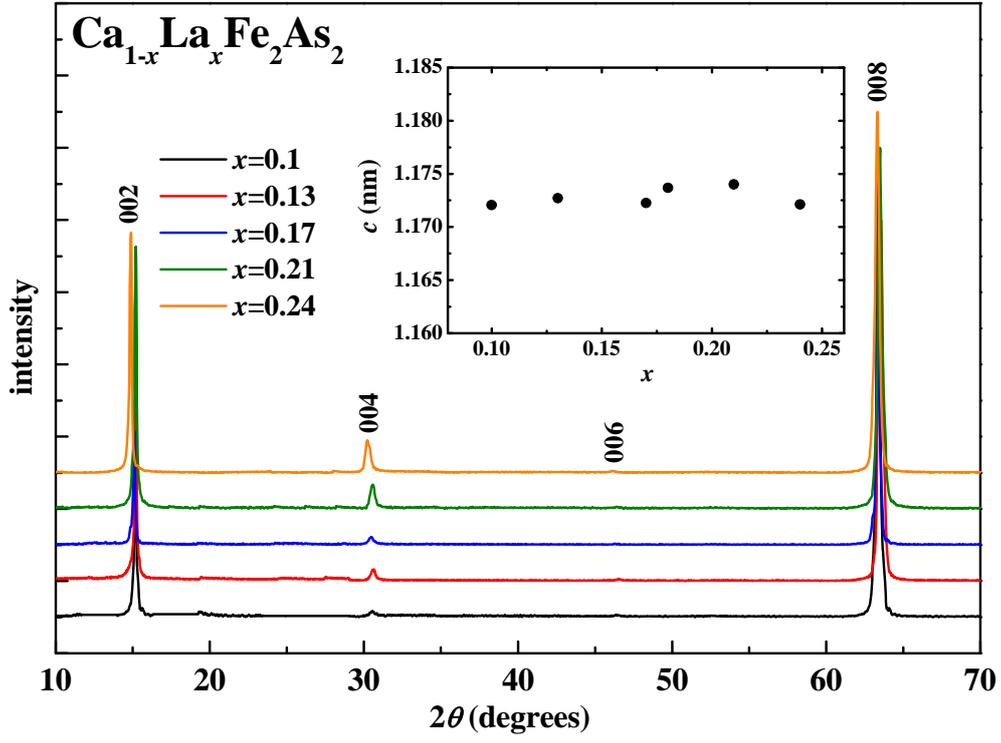

Figure 2

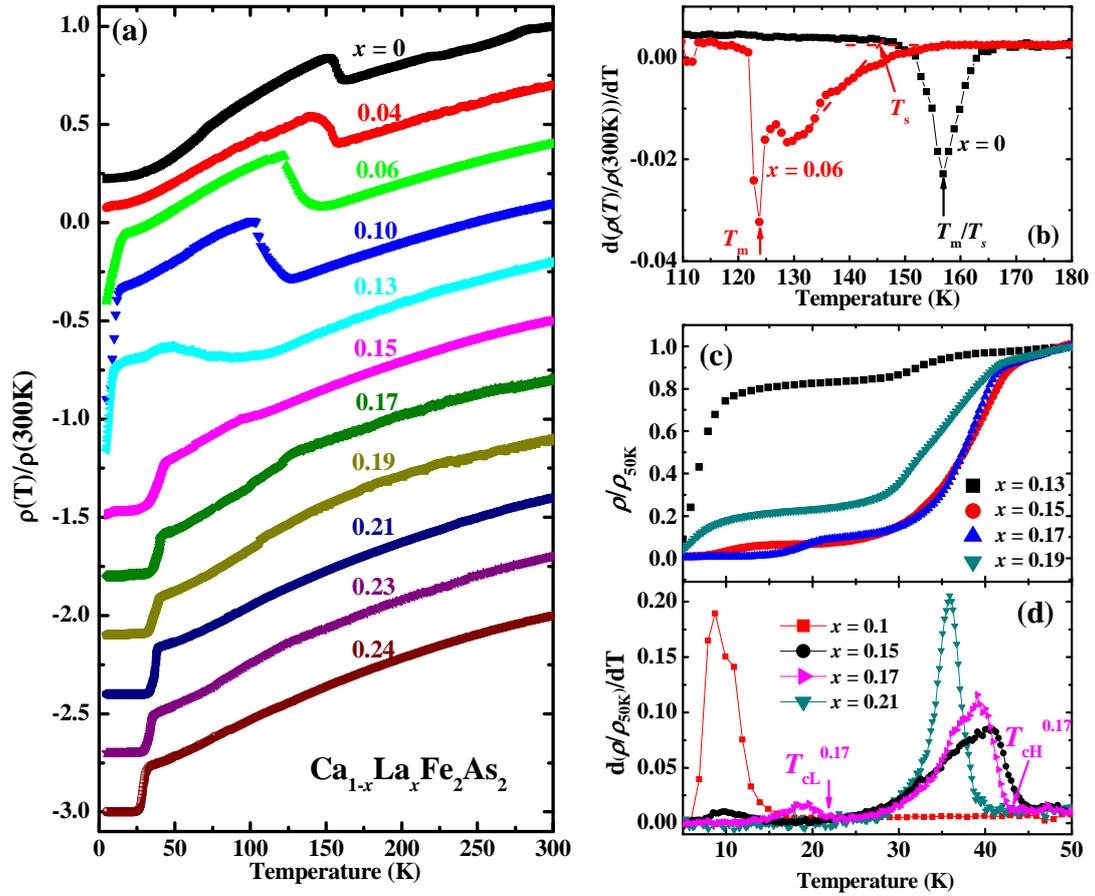

**Figure 3**

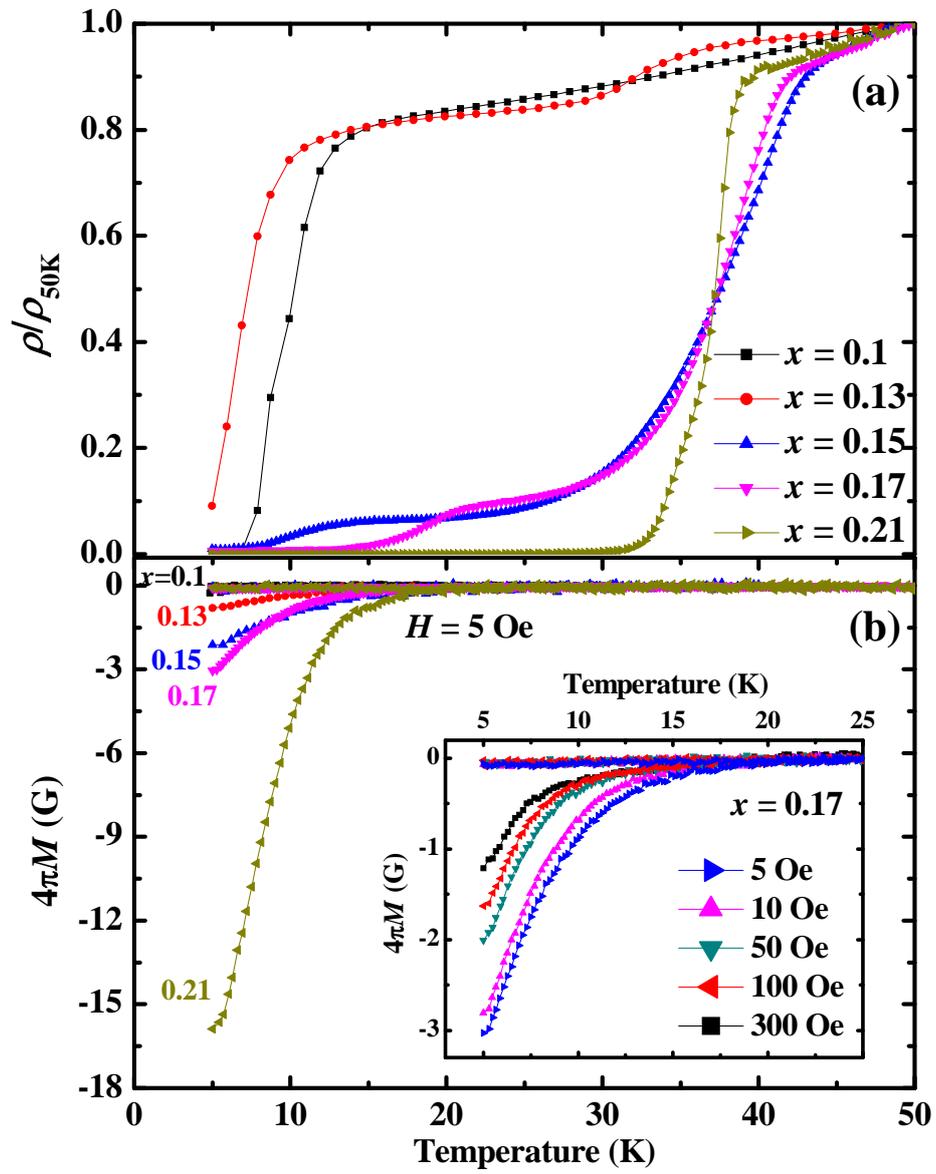

**Figure 4**

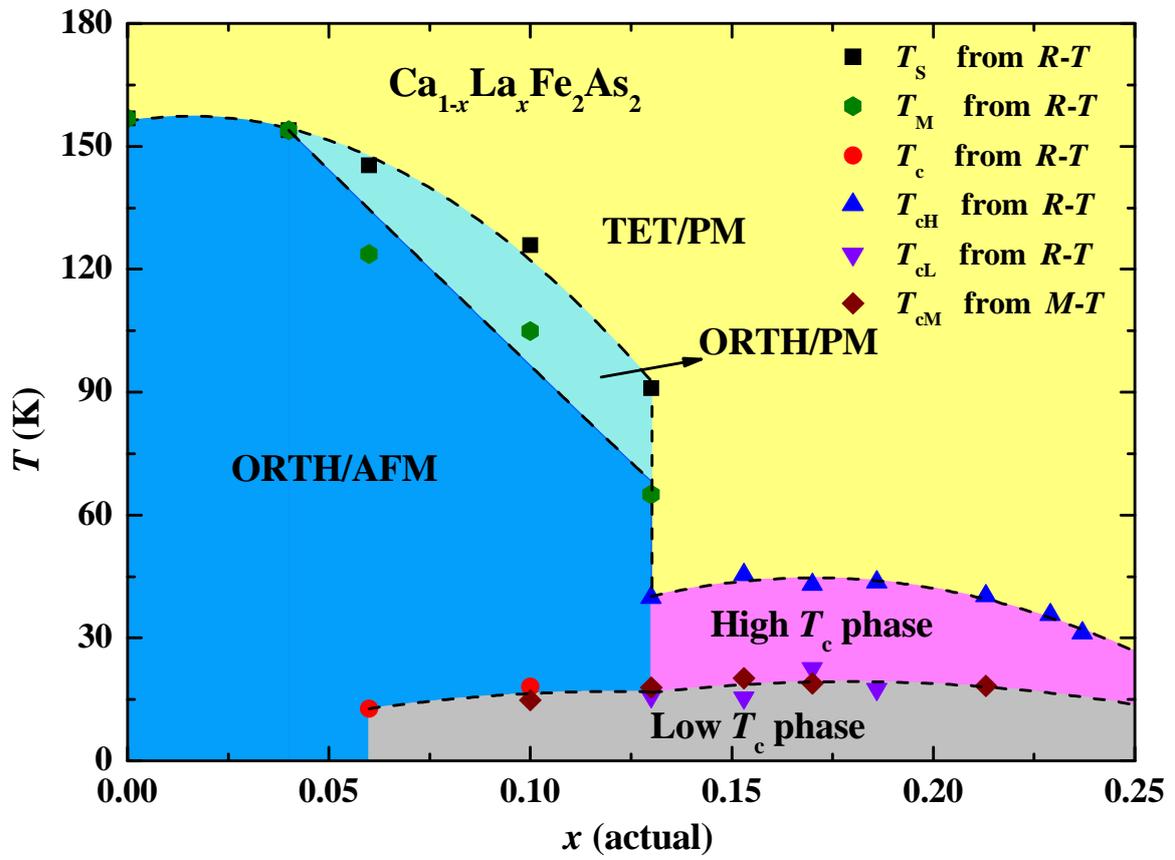

Figure 5

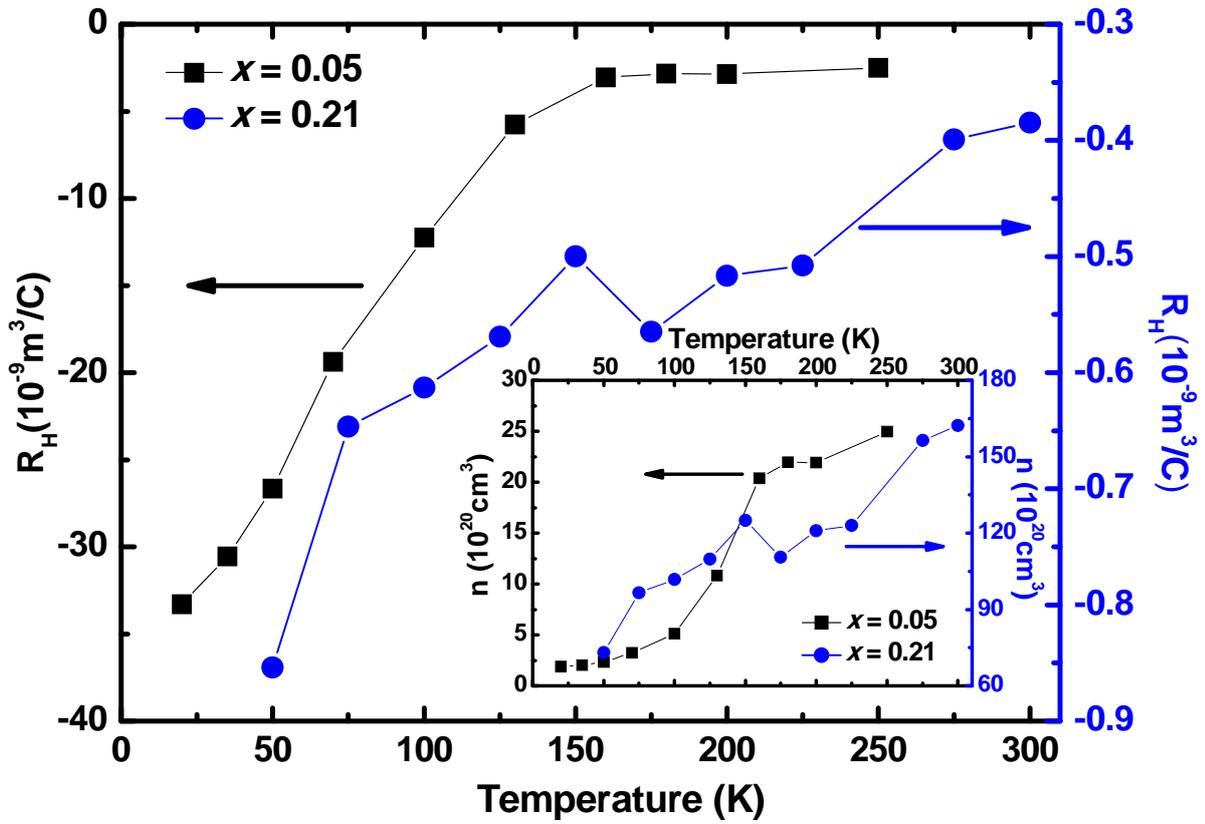

**Figure 6**

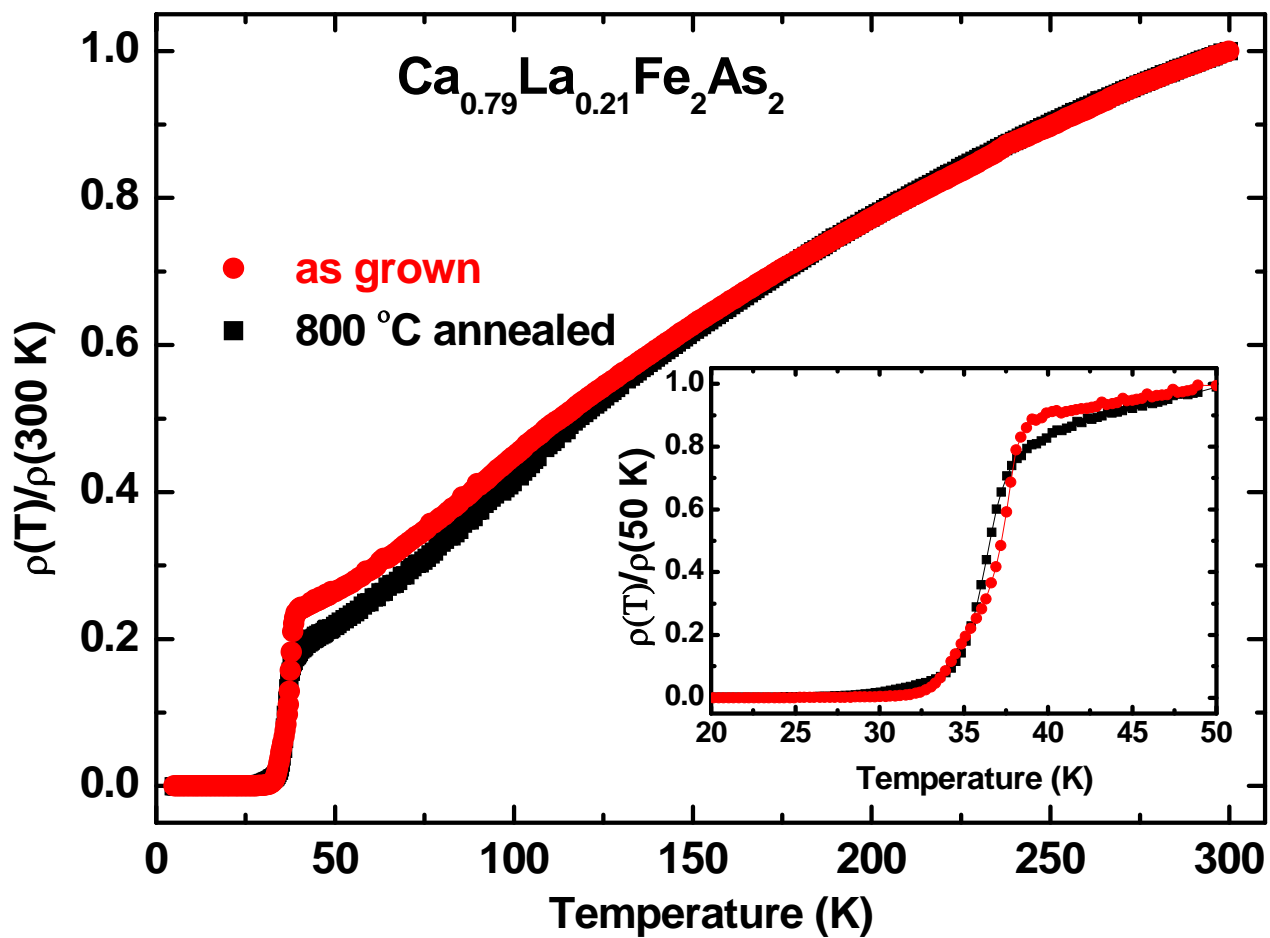

**Figure 7**

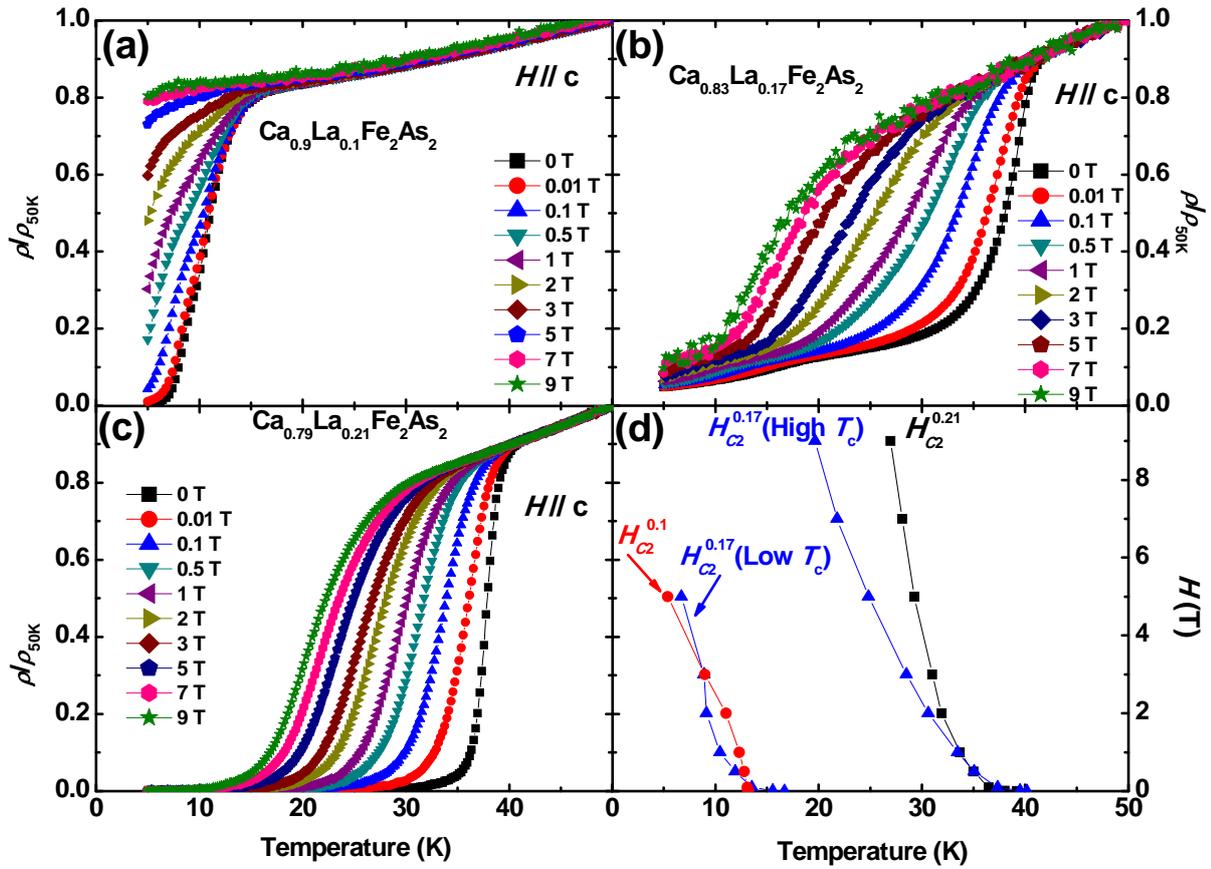

**Figure 8**